\documentclass{ws-p8-50x6-00}
\usepackage{rotating}

\begin{document}

\title{GENIUS and the Genius TF:\\ A New Observatory for WIMP Dark Matter 
and Neutrinoless Double Beta Decay}

\author{H.V.~Klapdor-Kleingrothaus\thanks{Spokesman of HEIDELBERG-MOSCOW and 
GENIUS collaborations; ~ E-mail: klapdor@gustav.mpi-hd.mpg.de}, 
B. Majorovits\thanks{Talk presented by B. Majorovits}}

\address{Max Planck Institut f\"ur Kernphysik, P.O. Box 103980,\\ 69029 
Heidelberg, Germany,\\ Home Page Heidelberg Non-Accelerator Particle 
Physics group: http://mpi-hd.mpg.de.non$\_$acc/}

\maketitle


\abstracts{The GENIUS proposal is described and some of it's physics 
potential is outlined. Also in the light of the contradictive results 
from the DAMA and CDMS experiments 
the Genius TF, a new experimental setup is proposed. 
The Genius TF could probe the DAMA evidence region using the WIMP 
nucleus recoil signal and WIMP annual modulation signature 
simultaneously. Besides that it can prove the long term feasibility 
of the detector technique to be implemented into the GENIUS setup and 
will in this sense be a first step towards the realization of the 
GENIUS experiment.}

\section{Introduction}
The topic of Dark Matter search has lately gained particular actuality by the 
results of the
DAMA \cite{damahere} and CDMS \cite{cdmshere} experiments. 
The DAMA collaboration claims to see positive evidence for WIMP
dark matter using the annual modulation signature, whereas the CDMS 
experiment seems to almost exclude fully the DAMA allowed cross-sections
for WIMP dark matter. It is, therefore, of utmost importance 
to independently test these results using both
experimental approaches: to look for the 
WIMP-nucleus recoil signal and for the annual
modulation effect. However, should the positive 
DAMA WIMP evidence be disproven, a large
step forward in terms of increasing the sensitivity of Dark Matter experiments 
is needed in order to obtain relevant data concerning WIMP dark matter.
In this article we shortly highlight the physics potential of the 
GENIUS project \cite{genius,GENIUS,geniusproposal} regarding WIMP Dark
Matter search, neutrinoless double beta  
decay and the real time observation of Solar neutrinos and we introduce the 
Genius TF \cite{gtfmpg,geniustf,diss}, 
a new experimental setup to probe the evidence region favoured by the
DAMA experiment \cite{damahere} and to test the prerequisites
neccessary to realize the Genius project.

\begin{figure}
\epsfxsize=25pc
\epsfbox{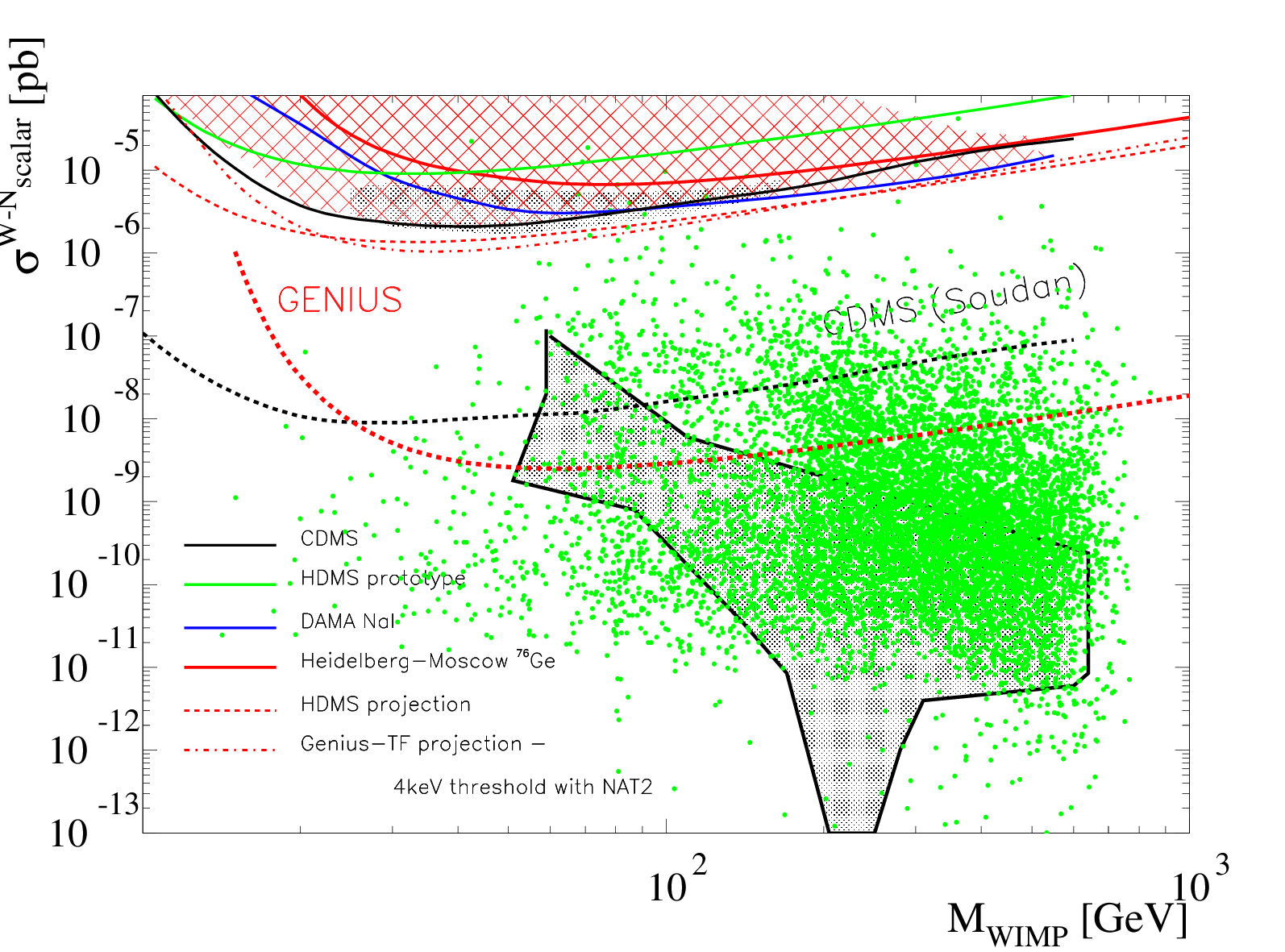}
\caption[]{Exclusion plot of the scalar 
WIMP-nucleon elastic scattering cross 
section as a function of the WIMP mass. Plotted are excluded areas from
the presently most sensitive direct detection experiments (hatched area
DAMA \cite{damaexcl}, CDMS \cite{cdmshere}, 
Heidelberg-Moscow \cite{WIMPS}, HDMS prototype \cite{hdms})
and some projections for experiments running or being presently under 
construction (HDMS, Genius TF \protect{\cite{gtfmpg,geniustf}}). 
The small shaded area represents the 2$\sigma$ evidence region from the
DAMA experiment \cite{damahere}. 
The extrapolated sensitivities of future
experiments (GENIUS \cite{geniusproposal}, CDMS at Soudan \cite{cdmshere}) 
are also shown. The scatter plot 
corresponds to predictions from theoretical considerations in
the MSSM with relaxed unification conditions \cite{klanew}. 
The large shaded area corresponds to calculations in the 
mSUGRA-inspired framework of the MSSM, with universality relations for the
parameters at GUT scale \cite{ellis}.
\label{dmall}}
\end{figure}

\section{The GENIUS experiment}
In order to achieve a dramatic step forward regarding background reduction, 
a new experimental technique is
needed. The GENIUS project uses the concept of
application of standard detection techniques while removing all 
dangerous contaminations from the direct vicinity of the detectors.

\subsection{The concept of the GENIUS experiment}
The GENIUS project \cite{genius,GENIUS,geniusproposal} is based on the
idea to operate 'naked' HPGe crystals 
directly in liquid nitrogen \cite{heusser2}.
The naked Germanium crystals are located in a huge nitrogen tank
(diameter 12-13\,m). This way
all dangerous
contaminations from the direct vicinity of the crystals are removed.
This has the great advantage that the liquid nitrogen which is very clean
with respect to radiopurity due to its production process (fractional
distillation), can act simultaneously as cooling medium and shield against 
external activities. The conceptual design of the experiment is
shown in figure \ref{genius_scheme}.

It has been shown that with this approach a
reduction of background by three to four orders of magnitude can be achieved
\cite{genius,GENIUS,geniusproposal,geniustf,diss}. The final reachable
background index with a 12~m diameter GENIUS tank is estimated to be
around $\sim 10^{-2}$ 
counts/(kg~keV~y) in the low-energy region below 50~keV. 
The sensitivity regarding WIMP Dark Matter search 
reachable with this background and a total detector mass of 100 kg of natural
Germanium can be seen in Fig. \ref{dmall}.

\begin{figure}
\epsfxsize=16.5pc
\rotatebox{-90}{\epsfbox{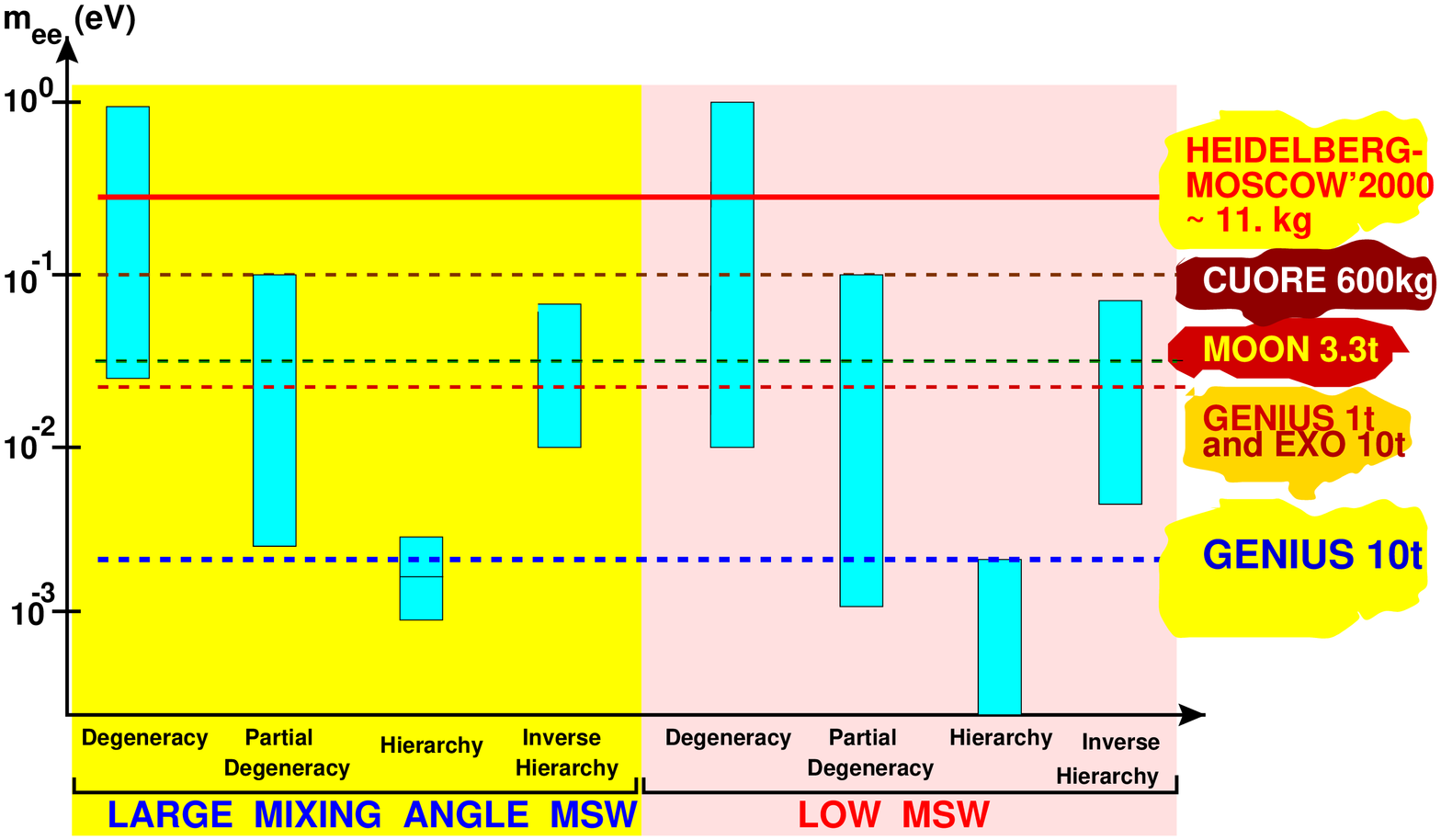}} 
\caption[]{Summary of values for 
$\langle$m$_{\nu}\rangle$ (here denoted as m$_{ee}$) expected from
present results of neutrino oscillation experiments in the 
different neutrino mass scenario schemes (from \cite{heinrichosc}).
The expectations are compared with the recent neutrino mass limits 
{\it obtained} from the Heidelberg--Moscow \cite{hdmo}
experiment as well as the expected 
sensitivities for the CUORE \cite{cuore}, MOON 
\cite{moon} proposals 
and the 1 ton and 10 ton proposal of GENIUS \cite{genius,geniusproposal}
\label{exclusion_intro} 
}
\end{figure}

With a background of $\sim$ 0.1 counts/(t keV y), which can be reached with 
this device in the energy region around the  
Q-value of the neutrinoless double beta decay of $^{76}$Ge at
2038.5~keV, and
with 1 tonne of enriched $^{76}$Ge, 
GENIUS would be sensitive to an effective Majorana neutrino mass 
down to $\sim$0.01~eV. This will allow already to test many
different neutrino mass scenarios (see figure \ref{exclusion_intro}).
If the sensitivity of GENIUS would be increased to a level of 
$\langle$m$_{\nu}\rangle \sim$~0.001~eV (using 10 tonnes of
$^{76}$Ge), this would allow to test {\it all} neutrino mass scenarios 
allowed by present neutrino oscillation experiments - except for one,
the (not favoured) hierarchical LOW solution. For a detailed
discussion see \cite{heinrichosc,16,17,18}.

With a background index of 10$^{-3}$counts/(kg~keV~y) in the energy region 
below 200~keV, which can be reached with a tank size of 13~m diameter
and some improved shielding\cite{baudissolar,20}, GENIUS (10 tons)
would be able to see  
the full solar pp neutrino spectrum in real time
\cite{baudissolar,geniusproposal,20},
with a count rate being a factor of 30-60 larger than a 20 tonnes LENS 
detector, and with a threshold of 11~keV or 19~keV.

\begin{figure}[t]
\epsfysize=20pc
\hspace*{2.75cm}
\epsfbox{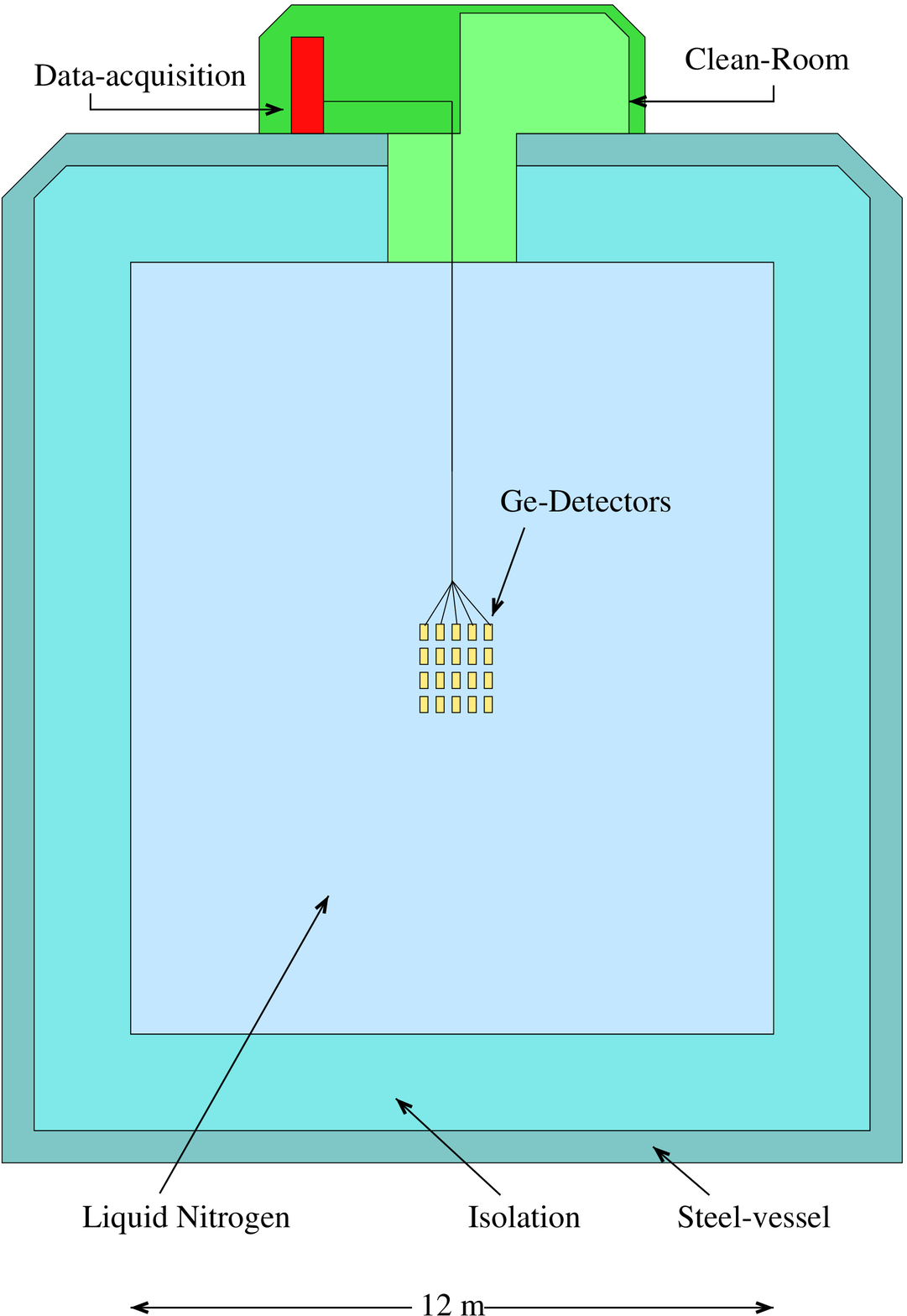}
\caption[]{\label{genius_scheme} Schematic view of the GENIUS project. 
An array 
of 100~kg of natural HPGe detectors for the WIMP dark matter search 
(first step) 
or 1 to 10~t of
enriched $^{76}$Ge for the double beta decay search (final setup) is hanging 
on a support
structure in the middle of the tank immersed in liquid nitrogen. The size of
the nitrogen shield would be 12 meters in diameter at least. On top of the tank
a special low level clean room and the room for the electronics and data 
acquisition will be placed. }
\end{figure}

\subsection{Tritium production in HPGe at sea level}
As evident from previous considerations of the expected background
\cite{geniusproposal,NIM}, great care has to be taken about the
cosmogenic isotopes produced inside the HPGe crystals at sea level. 
However, 
with an additional shield against the hard component of cosmic rays during
fabrication, the good sensitivity for dark matter can be maintained.

Especially the production of $^{68}$Ge from the isotope
$^{70}$Ge can affect the sensitivity due to the 10.37~keV X-ray emitted 
by the decay of  $^{68}$Ge to $^{68}$Ga.
In the main reaction leading to $^{68}$Ge enhancement
tritium is produced through the process
$^{70}$Ge(n,t)$^{68}$Ge. Tritium has a half life of 
12.35 years and can thus not be deactivated within a reasonable time.
$^3$H is a $\beta$ emitter with a Q-value of 18.6~keV. 

The cosmogenic production rate of $^3$H in natural germanium 
has been estimated  through simulations in \cite{juan,avignone} using the 
cosmic neutron fluxes cited in \cite{lal,hess}. 
For natural germanium it is estimated 
to be less than $\sim$ 200 atoms per day and kg material. 
Using this upper limit for tritium production at sea level
with an overall fabrication time of ten days this would
mean a tritium abundance of $\sim$ 2000 atoms per kilogram material.
With the half life of 12.3 years this results in a decay rate of
$\sim$3.6 $\mu$Bq/kg equivalent to $\sim$113 decays per year (this is in
very good agreement with the result in \cite{zdesenko}).
Even assuming an energy threshold of 12~keV and taking into account the 
spectral shape of tritium decay this yields
an event rate of approximately 2 counts/(kg~keV~y) in the energy
region between 
12~keV and 19~keV, which is by two orders of magnitude above the 
allowed count rate required for GENIUS as a dark matter detector.

This consideration drastically shows the importance of proper planning
of the crystal production and transportation.
To avoid major problems with cosmogenic isotopes it is
therefore essential to minimize the exposure of the crystals to cosmic rays at
sea level.

It is therefore planned to shield the detector material
during the complete 78 hours of production after the zone refining and
approximately one week of transportation periods using 
concrete with a thickness equivalent to $\sim$ 5~mwe. 
Heavy concrete can be produced with a density up to 5.9~g/cm$^3$. Thus an
additional heavy concrete layer of 1~m could act as a shield of roughly 5~mwe.
This reduces the hard nucleonic component mainly responsible for the 
cosmogenic isotope production by up to two orders of magnitude 
\cite{heusser2}. A further increase of shielding strength does not seem to be
reasonable since the cosmogenic production through fast muons 
which is by approximately two orders of magnitude less than through the 
hadronic component can not be shielded whatsoever.

With such a protection a reduction of the tritium 
production by a factor of $\sim$ 30 (see figures 2 and 3 in
\cite{heusser2}) can be obtained.
In this way the final background from tritium in the energy interval between 
12~keV and 19~keV would be $\sim$1.6$\times10^{-2}$ counts/(kg~keV~y)
without additional transportation and $\sim$5.6$\times10^{-2}$
counts/(kg~keV~y) with a 
week of transport from the fabrication site to the site of the experiment.

\section{The Genius-TF}

It has been shown\cite{bargein,diss} that with a setup
using a conventional shield, a sensitivity can be reached which allows for a
test of the DAMA evidence region within a short time period.
With an active mass of the detector of approximately
40 kg, projected for the Genius-TF (see
Fig. \ref{geniustf_scheme}), in which the background index will be  
maintained, not only the WIMP-nucleon recoil spectrum, but also the
expected signature of WIMP dark matter in form of the 
annual modulation signal could be tested within a reasonable time window
with a sensitivity probing the full DAMA evidence region
\cite{gtfmpg,geniustf,kla5}.

The Genius Test Facility could test the following points:
the long term stability of 'naked' HPGe-detectors in liquid nitrogen, the
possibility of constructing a feasible holder system and in addition
the DAMA evidence contour, through testing the expected signal and signature.
The GENIUS-TF will also be a suitable place to develop and 
test the electronics needed for the GENIUS experiment. The data
acquisition system should be based on a modular structure being capable of
taking data from up to 300 detectors simultaneously.

\subsection{The Test Facility}
The concept of the GENIUS proposal has the great advantage that no 
individual cryostat system is needed.
Instead the HPGe crystals are surrounded by liquid nitrogen of much
higher radiopurity which in addition provides ideal cooling and
shielding against external radiation. This opens the new research
potentials for the Genius project \cite{genius,GENIUS}.

It is proposed to install a setup with up to fourteen detectors 
on a small scale in order to be sensitive in the range
of the DAMA 
result \cite{damahere} on a short time scale and to prove the long term
stability of the new detector concept.

\begin{figure}[t]
\epsfysize=15pc
\hspace{2.75cm}
\epsfbox{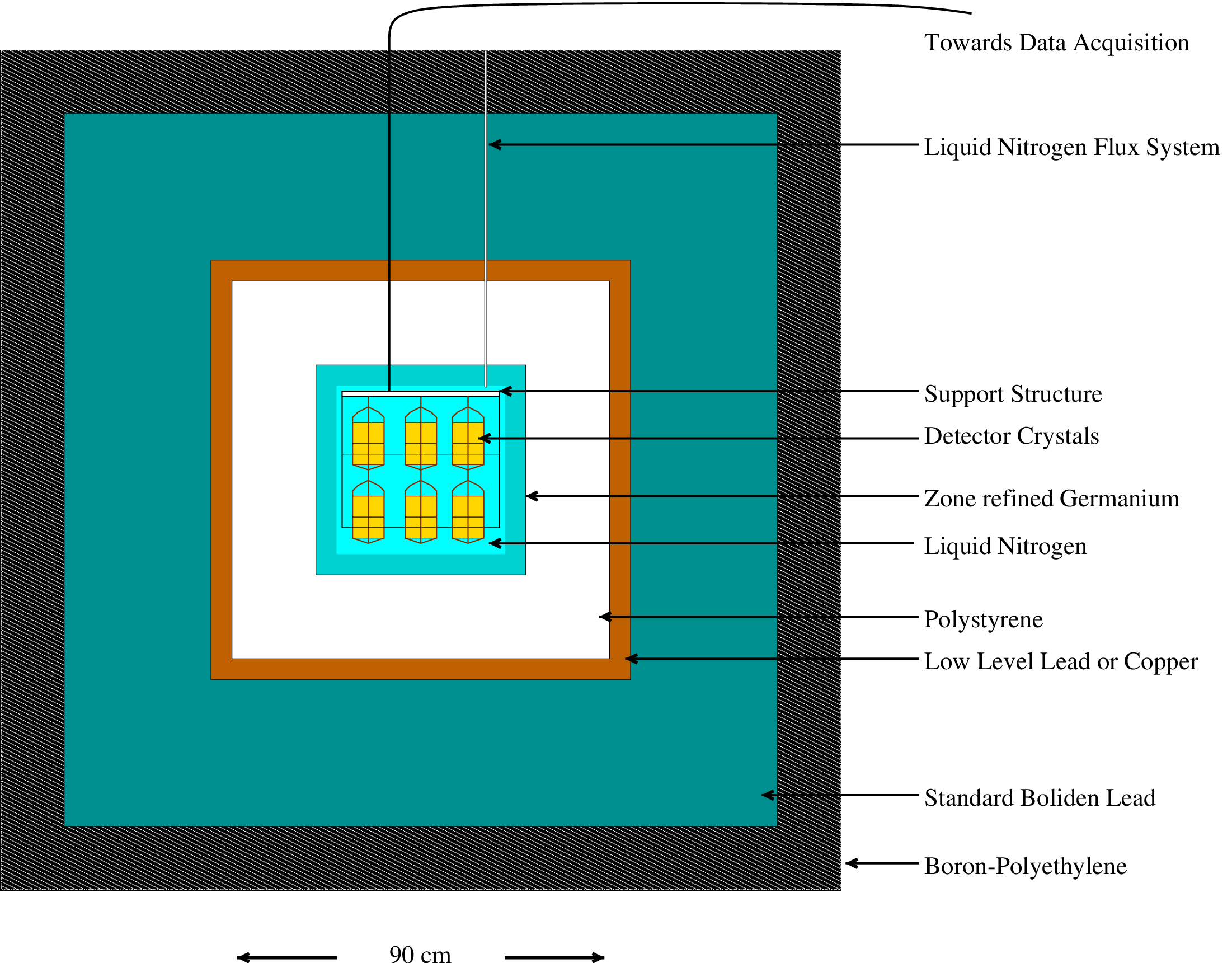}
\caption[]{\label{geniustf_scheme} 
Conceptual design of the Genius TF. Up
to 14 detectors will be housed in the inner detector chamber, filled with
liquid nitrogen. As a first shield 5 cm of zone refined Germanium will be used.
Behind the 20 cm of polystyrene isolation another 35 cm of low level lead
and a 15 cm borated polyethylene shield will complete the setup.}
\end{figure}

The design is shown in figure
\ref{geniustf_scheme}. 
It is based on a dewar made from low-activity polystyrene
and on a shield of zone refined germanium bricks inside the dewar and
low activity lead outside the dewar. A layer of boron loaded
polyethylene plates for suppression of neutron-induced background
completes the shield.

340 kg of zone-refined
high-purity Germanium bricks will serve as the inner 
layer to shield the 'naked' HPGe detector against the less radio-pure
polystyrene.  Also the first 5cm layer outside the
polystyrene-dewar needs to be of extreme radiopurity. The same type of 
copper as
installed in the Heidelberg-Moscow experiment, and/or some complementary
low-level lead could be used.
To shield the external $\gamma$ rays (natural radioactivity from the
surroundings) an overall lead layer of approximately 35 cm is needed. 

Using this concept an inner detector chamber of
40~cm$\times$40~cm$\times$40~cm 
would be sufficient to house up to seven HPGe-detectors in one layer or
14 detectors in two layers. This will 
allow for the development and test of a holder system for the same
amount of crystals.

The overall dimension of the experiment will be 
1.8~m$\times$1.8~m$\times$1.8~m thus fitting in one of the buildings
of the Heidelberg-Moscow experiment which is used momentarily for material measurements.

The background considerations and simulations discussed in
\cite{geniustf}, which extend those of \cite{bargein} suggest
that a reduction of the background by a factor 
of $\sim$5-10 with respect to the Heidelberg-Moscow-Experiment can be
attained with the proposed setup.

Assuming a final target mass of 40~kg, an energy threshold of 12~keV and
a background index of 4 counts/(kg keV y)
corresponding to $\sim$ 0.01 counts/(kg keV d) in the energy region between 
12~keV and 100~keV the Genius TF would need a significance of
190~kg~y to see the claimed DAMA annual modulation with 95\% probability
and 90\%C.L. (see \cite{cebrian}). This corresponds to an overall measuring 
time of approximately five years which would correspond to the life time of
this experiment.

However,  the new detectors will have an energy threshold of 0.5~keV
(four detectors with 2.5\,kg weight each, and this threshold have been 
produced already)
thus allowing for the use of the experimental
spectrum in the energy range between the threshold and the X-ray peaks seen
from the cosmogenically produced isotopes. This will significantly improve
the sensitivity of the Genius TF on the annual modulation effect.

In the energy region around the Q-value of the double beta decay of $^{76}$Ge
at 2038.5~keV a background index of $\sim7\times10^{-3}$counts/(kg keV y) 
could be reached, leading to a sensitivity for the effective Majorana 
neutrino mass down to 0.15~eV with 90\% C.L. \cite{gtfmpg,geniustf}

The construction of the setup
will be started in 2001. 
First results may be expected in the end of the year 2002.

\section{Conclusions}

The large physics potential of the GENIUS project regarding WIMP Dark
Matter search, neutrinoless double beta decay search and real time
observation of solar pp-neutrinos has been briefly outlined. 
We presented the Genius TF, a test facility for the GENIUS project,
whose construction started in early 2001.
The Genius TF can, according to Monte Carlo simulations, reach a
background of $\sim$ 2-4\,counts/(kg\,keV\,y) in the energy region
between 11\,keV and 100\,keV. Thus it could for the first 
time probe the DAMA evidence region using both, the WIMP-nuclear
recoil signal and the annual modulation signature.


\begin{thebibliography}{99}
\bibitem{damahere} P. Belli, these Proceedings and CDMS Collaboration 
  and R.~Bernabei et al. Phys. Lett. B \textbf{480}(2000)23
\bibitem{cdmshere} R. Gaitskell, these Proceedings, 
  R. Abusaidi et al., Nucl. Inst. and Meth. A \textbf{444}(2000)345
\bibitem{genius} H.V.~Klapdor-Kleingrothaus ~in Proceedings of the
  First International Conference on Particle Physics Beyond the
  Standard Model, {\it BEYOND THE DESERT 1997}, Castle Ringberg,
  Germany, 8-14 June 1997, edited by
  H.V.~Klapdor-Kleingrothaus~ and H.P\"as, IOP Bristol,
  1998, pp. 485--531, and Int. Journ. Mod. Phys. A {\bf 13}(1998)3953 
\bibitem{GENIUS} H.V.~Klapdor-Kleingrothaus, J. Hellmig und M. Hirsch,
  J. Phys. G \textbf{24}(1998)483
\bibitem{geniusproposal} H.V.~Klapdor-Kleingrothaus, L. Baudis,
  G. Heusser, B. Majorovits, H. P\"as, Proposal, MPI-H-V26-1999,
  August 1999, hep-ph/9910205 and in Proceedings of the Second
  International Conference on Particle Physics Beyond the Standard
  Model, {\it BEYOND THE DESERT
    1999}, Castle Ringberg, Germany, 6-12 June, 1999, ed. by 
  H.V.~Klapdor--Kleingrothaus, I. Krivosheina (IOP Bristol 2000),
  pp. 915-1024
\bibitem{gtfmpg}  H.V.~Klapdor-Kleingrothaus, L.~Baudis, A.~Dietz, 
  G.~Heusser, I.~Krivosheina, B. Majorovits, H.~Strecker , H.~Tu, et al., 
  Internal Report, Proposal MPI-H-V32-2000
\bibitem{geniustf} L. Baudis, A. Dietz, G. Heusser, B. Majorovits, 
  H. Strecker and H.V. Klapdor-Kleingrothaus, hep-ex/0012022, submitted
  for publication
\bibitem{diss} B. Majorovits, PhD thesis, University of Heidelberg, 2000
\bibitem{heusser2} G. Heusser, Ann. Rev. Nucl. Part. Sci. \textbf{45}(1995)543
\bibitem{NIM} L. Baudis, G. Heusser, H.V.~Klapdor-Kleingrothaus, B. Majorovits,
  Y. Ramachers, H. Strecker, Nucl. Inst. and Meth. A \textbf{426}(1999)425
\bibitem{damaexcl} R. Bernabei et al., Nucl. Phys. B (Proc. Suppl) 
  \textbf{70}(1998)79
\bibitem{WIMPS} HEIDELBERG-MOSCOW collaboration, L.~Baudis, 
  J.~Hellmig, G.~Heusser, H.V.~Klapdor-Kleingrothaus, 
  S~Kolb, B.~Majorovits, H.~P\"as, Y.~Ramachers, H.~Strecker, V.~Alexeev, 
  A.~Bakalyarov, A.~Balysh, S.T.~Belyaev, V.I.~Lebedev, S.~Zhoukov,
  Phys. Rev. D \textbf{59}(1998)022001 and Preprint hep-ex/9811045
\bibitem{hdms} L. Baudis, A. Dietz, B.~Majorovits, F.~Schwamm, 
  H.~Strecker, H.V.~Klapdor-Kleingrothaus, Phys. Rev. D 
\textbf{63}(2000)022001 and astro-ph/0008339 
\bibitem{klanew} V. Bednyakov and H.V.~Klapdor-Kleingrothaus,
  hep-ph/0011233, Phys. Rev. D, in press, (2001)
\bibitem{ellis} J. Ellis, A. Ferstl, K.A. Olive, Phys. Lett. B
  \textbf{481}(2000)304, hep-ph/0007113
\bibitem{heinrichosc} H.V. Klapdor-Kleingrothaus, H. P\"as,
  Y.A. Smirnov, hep-ph/0003219, Phys. Rev. D, in press, (2001)
\bibitem{16} H.V. Klapdor-Kleingrothaus, H. P\"as, Yu. Smirnov,
  submitted for publ. 
\bibitem{17} H.V. Klapdor-Kleingrothaus, H. P\"as, Comments in
  Nucl. and Part. Phys, in press, (2000) and physics/0006024
\bibitem{18} H.V. Klapdor-Kleingrothaus, in Proc. of NOON 2000 -
  Internat. Workshop on 
  'NEUTRINO OSCILLATIONS AND THEIR ORIGIN', Tokyo, Dec. 2000, World
  Scientific, Singapore, 2001 
\bibitem{hdmo} HEIDELBERG-MOSCOW Collaboration,
  Phys. Rev. Lett. \textbf{83}(1999)41
\bibitem{cuore} E. Fiorini et al., Phys. Rep \textbf{307}(1998)309 
\bibitem{moon} H. Eijiri et al., Phys.Rev.Lett. \textbf{85}(2000)2917,
  nucl-ex/9911008 
\bibitem{baudissolar} L. Baudis, H.V. Klapdor-Kleingrothaus, 
  Eur. Phys. J. A \textbf{5}(1999)441
\bibitem{20} H.V. Klapdor-Kleingrothaus, in Proc. of LowNu 2000 -
  Internat. Workshop on 
  'Low Energy Solar Neutrinos', Tokyo, Dec. 2000, World
  Scientific, Singapore, 2001 
\bibitem{juan} J. Collar, PhD thesis, University of South Carolina,
  1992
\bibitem{avignone} F.T. Avignone, et al., Nucl. Phys. B (Proc. Suppl.)
  \textbf{28}(1992)280
\bibitem{lal} D. Lal, B. Peter, {\it Cosmic Ray Produced Radioactivity 
    on the Earth}, Springer, Berlin-Heidelberg, 1967
\bibitem{hess} W.N. Hess, H.W. Patterson and R. Wallace, Phys. Rev.
  \textbf{116}(1959)449
\bibitem{zdesenko} O.A. Ponkratenko, V.I. Tretyak and Y.G. Zdesenko, 
  in Proc. of DARK98, 2nd International
  Conference on Dark Matter in Astro- and Particle Physics,
  Heidelberg, Germany, July 20-25, 1998, eds. H.V.~Klapdor-Kleingrothaus,
  L. Baudis, IoP, Bristol, 1999 
\bibitem{bargein} B. Majorovits, L.~Baudis, G.~Heusser, 
  H.~Strecker, H.V.~Klapdor-Kleingrothaus, Nucl. Inst. and Meth. A
  \textbf{455}(2000)371 
\bibitem{kla5} Homepage of the Non-Accelerator Particle Physics
  group, Max Planck Institut f\"ur
  Kernphysik, Heidelberg at http://www.mpi-hd.mpg.de/non\_acc   
\bibitem{cebrian} S. Cebrian et al., Astropart.Phys. \textbf{14}(2001)339 
and hep-ph/9912394
\end{thebibliography}
\end{document}